\documentclass[conference,a4paper]{IEEEtran}
\usepackage{listings}
\usepackage{qtree}
\usepackage{amsmath}
\usepackage{forest}
\usepackage{color}
\usepackage{array}
\usepackage{graphicx}
\usepackage{float}
\usepackage{xspace}
\usepackage{eurosym}
\usepackage{commath}
\usepackage[labelfont=bf]{caption}

\newcommand{\BigO}[1]{$O\left(#1\right)$\xspace}

\newcommand{\geoplus}{{geohash}\textsuperscript{+}\xspace}
\newcommand{\Geoplus}{{Geohash}\textsuperscript{+}\xspace}

\graphicspath{ {graphics/} }
\definecolor{dkgreen}{rgb}{0,0.6,0}
\definecolor{gray}{rgb}{0.5,0.5,0.5}
\definecolor{mauve}{rgb}{0.58,0,0.82}

\usepackage[bookmarks=false]{hyperref}

\newcommand\copyrighttext{%
	\footnotesize \textcopyright © 2020 IEEE.  Personal use of this material is permitted.  Permission from IEEE must be obtained for all other uses, in any current or future media, including reprinting/republishing this material for advertising or promotional purposes, creating new collective works, for resale or redistribution to servers or lists, or reuse of any copyrighted component of this work in other works.}
\newcommand\copyrightnotice{%
	\begin{tikzpicture}[remember picture,overlay]
	\node[anchor=south,yshift=10pt] at (current page.south) {\fbox{\parbox{\dimexpr\textwidth-\fboxsep-\fboxrule\relax}{\copyrighttext}}};
	\end{tikzpicture}%
}

\title{A rapidly updating stratified mix-adjusted median property price index model}

\author{
	\IEEEauthorblockN{Robert Miller\IEEEauthorrefmark{1} and
		Phil Maguire\IEEEauthorrefmark{2}}
	\IEEEauthorblockA{Dept. of Computer Science,
		National University of Ireland, Maynooth,\\
		Kildare, Ireland. \\
		Email: \IEEEauthorrefmark{1}robert.miller@mu.ie,
		\IEEEauthorrefmark{2}phil.maguire@mu.ie}
}

\begin{document}
\maketitle
\copyrightnotice

\begin{abstract}
	
	Homeowners, first-time buyers, banks, governments and construction companies are highly interested in following the state of the property market. Currently, property price indexes are published several months out of date and hence do not offer the up-to-date information which housing market stakeholders need in order to make informed decisions. In this article, we present an updated version of a central-price tendency based property price index which uses geospatial property data and stratification in order to compare similar houses. The expansion of the algorithm to include additional parameters owing to a new data structure implementation and a richer dataset allows for the construction of a far smoother and more robust index than the original algorithm produced.
	
\end{abstract}

\section{Introduction}

	House price indexes provide vital information to the political, financial and sales markets, affecting the operation and services of lending institutions greatly and influencing important governmental decisions \cite{Diewert2015}. As one of the largest asset classes, house prices can even offer insight regarding the overall state of the economy of a nation \cite{case_shiller_quigley_2001}. Property value trends can predict near-future inflation or deflation and also have a considerable effect on the gross domestic product and the financial markets \cite{FORNI20031243,Gupta2013}.
	
	There are a multitude of stakeholders interested in the development and availability of an algorithm which can offer an accurate picture of the current state of the housing market, including home buyers, construction companies, governments, banks and homeowners  \cite{Maguire2016,Plakandaras2014}.

	Due to the recent global financial crisis, house price indexes and forecasting models play a more crucial role than ever. The key to providing a more robust and up-to-date overview of the housing market lies in machine learning and statistical analysis on set of big data \cite{Hernando}. The primary aim is the improvement of currently popular algorithms for calculating and forecasting price changes, while making such indexes faster to compute and more regularly updated. Such advances could potentially play a key role in identifying price bubbles and preventing future collapses in the housing market \cite{Jadevicius2015,Klotz2016}.

	Hedging against market risk has been shown to be potentially beneficial to all stakeholders, however, it relies on having up-to-date and reliable price change information which is generally not publicly available \cite{Hernando,Englund2002}. This restricts the possibility of this tool becoming a mainstream option to homeowners and small businesses.
	
	In this article, we will expand upon previous work by \cite{Maguire2016} on a stratified, mix-adjusted median property price model by applying that algorithm to a larger and richer dataset of property listings and explore the enhancements in smoothness offered by evolving the original algorithm enabled by the use of a new data structure \cite{GeoTree}.
	
\section{Property price index models}
	In this section we will detail the three main classes of existing property price indexes. These consist of the \textit{hedonic regression}, \textit{repeat-sales} and \textit{central-price tendency} methods.
	
	\subsection{Hedonic Regression}
	 	Hedonic regression \cite{Kain1970} is a method which considers all of the characteristics of a house (eg. bedrooms, bathrooms, land size, location etc.) and calculates how much weight each of these attributes have in relation to the overall price of the house. While it has been shown to be the most robust measure in general by \cite{Goh2012}, outperforming the repeat-sales and mix-adjusted median methods, it requires a vast amount of detailed data and the interpretation of an experienced statistician in order to produce a result \cite{Maguire2016,Bourassa2006}. 
	 	
	 	As hedonic regression rests on the assumption that the price of a property can be broken down into its integral attributes, the algorithm in theory should consider every possible characteristic of the house. However, it would be impractical to obtain all of this information. As a result, specifying a complete set of regressors is extremely difficult \cite{CaseComparison}. 
	 	
	 	The great number of free parameters which require tuning in hedonic regression also leads to a high chance of overfitting the model \cite{Maguire2016}.
	
	\subsection{Repeat-sales}
		The repeat-sales method \cite{Bailey1963} is the most commonly used method of reporting housing sales in the United States and uses repeated sales of the same property over long periods of time to calculate change. An enhanced, weighted version of this algorithm was explored by \cite{NBERw2393}. The advantage of this method comes in the simplicity of constructing and understanding the index; historical sales of the same property are compared with each other and thus the attributes of each house need not be known nor considered. The trade-off for this simplicity comes at the cost of requiring enormous amounts of data stretched across long periods of time \cite{DEVRIES2009}. 
	
		It has also been theorised that the sample of repeat sales is not representative of the housing market as a whole. For example, in a study by \cite{Jansen2008}, only 7\% of detached homes were resold in the study period, while 30\% of apartments had multiple sales in the same dataset. It is argued that this phenomenon occurs due to the 'starter home hypothesis': houses which are cheaper and in worse condition generally sell more frequently due to young homeowners upgrading \cite{Jansen2008,Costello2002,DORSEY201075}. This leads to over-representation of inexpensive and poorer quality property in the repeat-sales method. Cheap houses are also sometimes purchased for renovation or are sold quickly if the homeowner becomes unsatisfied with them, which contributes to this selection bias \cite{Jansen2008}. Furthermore, newly constructed houses are under-represented in the repeat-sales model as a brand new property cannot be a repeat sale unless it is immediately sold on to a second buyer \cite{Costello2002}.
		
		As a result of the low number of repeat transactions, an overwhelming amount of data is discarded \cite{Dombrow1997}. This leads to great inefficiency of the index and its use of the data available to it. In the commonly used repeat-sales algorithm by \cite{NBERw2393}, almost 96\% of the property transactions are disregarded due to incompatibility with the method \cite{CaseComparison}.
	
	\subsection{Central Price Tendency}
	
		Central-price tendency models have been explored as an alternative to the more commonly used methods detailed previously. The model relies on the principle that large sets of clustered data tend to exhibit a noise-cancelling effect and result in a stable, smooth output \cite{Maguire2016}. Furthermore, central price tendency models offer a greater level of simplicity than the highly-theoretical hedonic regression model. When compared to the repeat sales method, central tendency models offer more efficient use of their dataset, both in the sense of quantity and time period spread \cite{Maguire2016,Prasad2008}. 
		
		According to a study of house price index models by \cite{Goh2012}, the central-tendency method employed by \cite{Prasad2008} significantly outperforms the repeat-sales method despite utilising much smaller dataset. However, the method is criticised as it does not consider the constituent properties of a house and is thus more prone to inaccurate fluctuations due to a differing mix of sample properties between time periods \cite{Goh2012}. For this reason, \cite{Goh2012} finds that the hedonic regression model still outperforms the mix-adjusted median model used by \cite{Prasad2008}. Despite this, the simplicity and data utilisation that the method offers deserve credit were argued to justify these drawbacks \cite{Prasad2008,Goh2012}.
		
		An enhancement to the mix-adjusted median algorithm by \cite{Prasad2008} was later shown to outperform the robustness of the hedonic regression model used by the Irish Central Statistics Office \cite{Maguire2016,OHanlon2011}. The primary drawback of this algorithm was long execution time and high algorithmic complexity due to brute-force geospatial search, limiting the algorithm from being further expanded, both in terms of algorithmic features and the size of the dataset \cite{GeoTree}.
		
		\subsection{Improvement Attempts}
		
		With an aim to overcome the issue of algorithmic complexity in the method described by \cite{Maguire2016}, a niche data structure was designed primarily for the purpose of greatly speeding up the geospatial proximity search with the aim of sacrificing minimal algorithmic precision. The \textit{GeoTree} offers a substantial performance improvement when applied to the original algorithm while producing an almost identical index \cite{GeoTree}. Through application of the GeoTree, the restrictions on the original algorithm have been lifted and we can now explore the performance of an evolved implementation of the algorithm on a richer, alternative dataset while introducing further parameters.
		
		\section{Case Study: MyHome Property Listing Data}

		MyHome \cite{MyHome} are a major player in property sale listings in Ireland. With data on property asking prices being collected since 2011, MyHome have a rich database of detailed data regarding houses which have been listed for sale. MyHome have provided access to their dataset for the purposes of this research.
		
		\subsection{Dataset Overview}
		
		The data provided by MyHome includes verified GPS co-ordinates, the number of bedrooms, the type of dwelling and further information for most of its listings. It is important to note, however, that this dataset consists of asking prices, rather than the sale prices featured in the less detailed Irish Property Price Register Data (used in the original algorithm) \cite{Maguire2016}.
		
		The dataset consists of a total of 718,351 property listing records over the period February 2011 to March 2019 (inclusive). This results in 7,330 mean listings per month (with a standard deviation of 1,689), however, this raw data requires some filtering for errors and outliers.
		
		\subsection{Data Filtration}
		
		As with the majority of human collected data, some pruning must be done to the MyHome dataset in order to remove outliers and erroneous data. Firstly, not all transactions in the dataset include verified GPS co-ordinates or include data on the number of bedrooms. These records will be instantly discarded for the purpose of the enhanced version of the algorithm. They account for 16.5\% of the dataset. Furthermore, any property listed with greater than six bedrooms will not be considered. These properties are not representative of a standard house on the market as the number of such listings amounts to just 1\% of the entire dataset.
		
		Any data entries which do not include an asking price cannot be used for house price index calculation and must be excluded. Such records amount to 3.6\% of the dataset. Additionally, asking price records which have a price of less than \EUR{10,000} or more than \EUR{1,000,000} are also excluded, as these generally consist of data entry errors (eg. wrong number of zeroes in user-entered asking price), abandoned or dilapidated properties in listings below the lower bound and mansions or commercial property in the entries exceeding the upper bound. Properties which meet these exclusion criteria based on their price amount to only 2\% of the dataset and thus are not representative of the market overall.
		
		In summation, 77\% of the dataset survives the pruning process. This leaves us with 5,646 filtered mean listings per month. 
		
		\subsection{Comparison with PPR Dataset}
		
		The mean number of filtered monthly listings available in our dataset represents a 157\% increase on the 2,200 mean monthly records used in the original algorithm's index computation \cite{Maguire2016}. Furthermore, the dataset in question is significantly more precise and accurate than the PPR dataset, owing to the ability to more effectively prune the dataset. The PPR dataset consists of address data entered by hand from written documents and does not use the Irish postcode system, meaning that addresses are often vague or ambiguous. This results in some erroneous data being factored into the model computation as there is no effective way to prune this data \cite{Maguire2016}. The MyHome dataset has been filtered to include verified addresses only, as described previously.
		
		The PPR dataset has no information on the number of bedrooms or any key characteristics of the property. This can result in dilapidated properties, apartment blocks, inherited properties (which have an inaccurate sale value which is used for taxation purposes) and mansions mistakenly being counted as houses \cite{Maguire2016}. Our dataset consists of only single properties and the filtration process described previously greatly reduces the number of such unrepresentative samples making their way into the index calculation.
		
		The "sparse and frugal" PPR dataset was capable of outperforming the CSO's hedonic regression model with a mix-adjusted median model \cite{Maguire2016}. With the larger, richer and more well-pruned MyHome dataset, further algorithmic enhancements to this model are possible.
		
		\section{Performance Measures}
		
		Property prices are generally assumed to change in a smooth, calm manner over time \cite{mcmillenhousepricesmooth} \cite{clapphousepricesmooth}. According to \cite{Maguire2016}, the smoothest index is, in practice, the most robust index. As a result of this, smoothness is considered to be one of the strong indicators of reliability for an index. However, the 'smoothness' of a time series is not well defined nor immediately intuitive to measure mathematically.
		
		The standard deviation of the time series will offer some insight into the spread of the index around the mean index value. A high standard deviation indicates that the index changes tend to be large in magnitude. While this is useful in investigating the "calmness" of the index (how dramatic its changes tend to be), it is not a reliable smoothness measure, as it is possible to have a very smooth graph with sizeable changes.
		
		The standard deviation of the differences is a much more reliable measure of smoothness. A high standard deviation of the differences indicates that there is a high degree of variance among the differences ie. the change from point to point is unpredictable and somewhat wild. A low value for this metric would indicate that the changes in the graph behave in a more calm manner. 
		
		Finally, we present a metric which we have defined, the \textit{mean spike magnitude $\mu_{\Delta{X}}$} (MSM) of a time series $X$. This is intended to measure the mean value of the contrast between changes each time the trend direction of the graph flips. In other words, it is designed to measure the average size of the 'spikes' in the graph. 
		
		Given $D_X = \{d_1, \dots, d_n\}$ is the set of differences in the time series $X$, we say that the pair $(d_i, d_{i+1})$ is a spike if $d_i$ and $d_{i+1}$ have different signs. Then $S_i = |d_{i+1} - d_i|$ is the spike magnitude of the spike $(d_i, d_{i+1})$. 
		
		The \textit{mean spike magnitude} of $X$ is defined as:
		
		{\large
			\[
			\mu_{\Delta{X}} = \frac{1}{\abs{S_X}}\sum\limits_{S \in S_X}{S^2}
			\]}
		
		\noindent
		where:
		\[
		 S_X = \{ S_1, S_2, ... , S_t \} \textrm{ is the set of all spike magnitudes of } X
		\]
		
		\section{Algorithmic Evolution}
		
		\subsection{Original Price Index Algorithm}
		
		The central price tendency algorithm introduced by \cite{Maguire2016} was designed around a key limitation; extremely frugal data. The only data available for each property was location, sale date and sale price. The core concept of the algorithm relies on using geographical proximity in order to match similar properties historically for the purpose of comparing sale prices. While this method is likely to match certain properties inaccurately, the key concept of central price tendency is that these mismatches should average out over large datasets and cancel noise.
		
		The first major component of the algorithm is the voting stage. The aim of this is to remove properties from the dataset which are geographically isolated. The index relies on matching historical property sales which are close in location to a property in question. As a result, isolated properties will perform poorly as it will not be possible to make sufficiently near property matches for them. 
		
		In order to filter out such properties, each property in the dataset gives one vote to its closest neighbour, or a certain, set number of nearest neighbours. Once all of these votes have been casted, the total number of votes per property is enumerated and a segment of properties with the lowest votes is removed. In the implementation of the algorithm used in \cite{Maguire2016}, this amounted to ten percent of the dataset.
		
		Once the voting stage of the algorithm is complete, the next major component is the stratification stage. This is the core of the algorithm and involves stratifying average property changes on a month by month comparative basis which then serve as multiple points of reference when computing the overall monthly change. The following is a detailed explanation of the original algorithm's implementation.
		
		First, take a particular month in the dataset which will serve as the stratification base, $m_b$. Then we iterate through each house sale record in $m_b$, represented by $h_{m_b}$. We must now find the nearest neighbour of $h_{m_b}$ in each preceding month in the dataset, through a proximity search. For each prior month $m_x$ to $m_b$, refer to the nearest neighbour in $m_x$ to $h_{m_b}$ in question as $h_{m_x}$. Now we are able to compute the change between the sale price of $h_{m_b}$ and the nearest sold neighbour to $h$ in each of the months $\{m_1, \dots, m_n\}$ as a ratio of $h_{m_b}$ to $h_{m_x}$ for $x \in \{1, \dots, n\}$. Once this is done for every property in $m_b$, we will have a scenario such that there is a catalogue of sale price ratios for every month prior to $m$ and thus we can look at the median price difference between $m$ and each historic month. 
		
		However, this is only stratification with one base, referred to as stage three in the original article \cite{Maguire2016}. We then expand the algorithm by using every month in the dataset as a stratification base. The result of this is that every month in the dataset now has price reference points to every month which preceded it and we can now use these reference points as a way to compare month to month. 
		
		Assume that $m_x$ and $m_{x+1}$ are consecutive months in the dataset and thus we have two sets of median ratios $\{r_x(m_1), \dots, r_x(m_{x-1})\}$ and $\{r_{x+1}(m_1), \dots, r_{x+1}(m_{x})\}$ where $r_{a}(m_y)$ represents the median property sale ratio between months $m_a$ and $m_y$ where $m_a$ is the chosen stratification base. In order to compute the property price index change from $m_x$ to $m_{x+1}$, we look at the difference between $r_x(m_i)$ and $r_{x+1}(m_i)$ for each $i \in {1, \dots, x-1}$ and take the mean of those differences. As such, we are not directly comparing each month, rather we are contrasting the relationship of both months in question to each historical month and taking an averaging of those comparisons.
		
		This results in a central price tendency based property index that outperformed the national Irish hedonic regression based index while using a far more frugal set of data to do so. 
		
		\subsection{GeoTree}
		
		The largest drawback of the original index lies in the computational complexity; it is extremely slow to run. This is due to the performance impact of requiring repeated search for neighbours to each data point. This limitation was responsible for preventing the algorithm scaling to larger datasets, more refined time periods and more regular updating. A custom data structure, the GeoTree, was developed in order to trade off a small amount of accuracy in return for the ability to retrieve a cluster of neighbours to any property in constant time \cite{GeoTree}. This data structure relies on representing the geographical location of properties as geohash strings.
		
		The GeoTree data structure functions by placing the geohash character by character into a tree-structure where each branch at each level represents an alphanumeric character. Under each branch of the tree there is also a list node which caches all of the property records which exist as an entry in that subtree, allowing the \BigO{1} retrieval of those records. The number of sequential characters in common from the start of a pair of geohashes puts a bound on the distance between those two geohashes. Thus, by traversing down the tree and querying the list nodes, the GeoTree can return a list of approximate nearest neighbours in \BigO{1} time \cite{GeoTree}.
		
		As can be seen in \cite[Table I]{GeoTree}, the performance improvement to the index offered by the GeoTree is profound and sacrifices very little in terms of precision, with the resulting indexes proving close to identical. This development allows the scope of the index algorithm to be widened, including the introduction of larger datasets with richer data, more frequent updating and the development of new algorithmic features, some of which will be explored in this article.
		
		\subsection{\Geoplus}
		
		Extended geohashes, which we will refer to as \geoplus, are geohashes which have been modified to encode additional information regarding the property at that location. Additional parameters are encoded by adding a character in front of the geohash. The value of the character at that position corresponds to the value of the parameter which that character represents. \autoref{fig:geo_plus} demonstrates the structure of a \geoplus with two additional parameters, $p_1$ and $p_2$.
		
		\begin{figure}[H]
			\[ \textrm{\geoplus: } \underbrace{p_1p_2}_{\textrm{+}} \underbrace{x_1 \dots x_n}_{\textrm{geohash}} \]
			
			\noindent
			\caption{\geoplus format}
			\label{fig:geo_plus}
		\end{figure}
		
		Any number of parameters can be prepended to the geohash. In the context of properties, this includes the number of bedrooms, the number of bathrooms, an indicator of the type of property (detached house, semi-detached house, apartment etc.), a parameter representing floor size ranges and any other attribute desired for comparison.
		
		Alternative applications of \geoplus could include a situation where a rapid survey of nearby live vehicles of a certain type is required. If we prepend a parameter to the geohash locations of vehicles representing that vehicle's type, eg: $1$ for cars, $2$ for vans, $3$ for motorcycles and so forth, we can use the GeoTree data structure to rapidly survey the SCBs around a particular vehicle, with separate SCBs generated for each type automatically.
		
		\subsection{GeoTree Performance with \geoplus}
		
		Due to the design of the GeoTree data structure, a \geoplus will be inserted into the tree in exactly the same manner as a regular geohash \cite{GeoTree}. If the original GeoTree had a height of $h$ for a dataset with $h$-length geohashes, then the GeoTree accepting that geohash extended to a \geoplus with $p$ additional parameters prepended should have a height of $h+p$. However, both of these are fixed, constant, user-specified parameters which are independent of the number of data points, and hence do not affect the constant-time performance of the GeoTree. 
		
		The major benefit of this design is that the ranged proximity search will interpret the additional parameters as regular geohash characters when constructing the common buckets upon insertion, and also when finding the SCB in any search, without introducing additional performance and complexity drawbacks.

		\subsection{Enhanced Price Index}
		
		In order to enhance our price index model, we prepend a parameter to the geohash of each property representing the number of bedrooms present within that property. As a result, when the GeoTree is performing the SCB computation, it will now only match properties which are both nearby and share the same number of bedrooms. This allows the index model to compare the price of properties which are more similar across the time series and thus should result in a smoother, more accurate measure of the change in prices over time.
		
		The technical implementation of this algorithmic enhancement is handled almost entirely by the GeoTree automatically, due to its design. As described previously, the GeoTree sees the additional parameter no differently to any other character in the geohash and due to its placement at the start of the geohash, the search space will be instantly narrowed to properties with matching number of bedrooms, $x$, by taking the $x$ branch in the tree at the first step of traversal.
		
		\section{Results}
		
		We ran the algorithm on the MyHome data without factoring any additional parameters as a control step. We then created a GeoTree with \geoplus entries consisting of the number of bedrooms in the house prepended to the geohash for the property.
				
		\subsection{Comparison of Time Series}
		
		\autoref{tab:index_stats} shows the performance metrics previously described applied to the algorithms discussed in this paper: Original PPR, PPR with GeoTree, MyHome without bedroom factoring and MyHome with bedroom factoring. While both the standard deviation of the differences and the MSM show that some smoothness is sacrificed by the GeoTree implementation of the PPR algorithm, the index running on MyHome's data without bedroom factoring approximately matches the smoothness of the original algorithm. Furthermore, when bedroom factoring is introduced, the algorithm produces by far the smoothest index, with the standard deviation of the differences being 26.2\% lower than the PPR (original) algorithm presented in \cite{Maguire2016}, while the MSM sits at 58.2\% lower.
		
\begin{figure*}
			\centering
			\includegraphics[width=\linewidth]{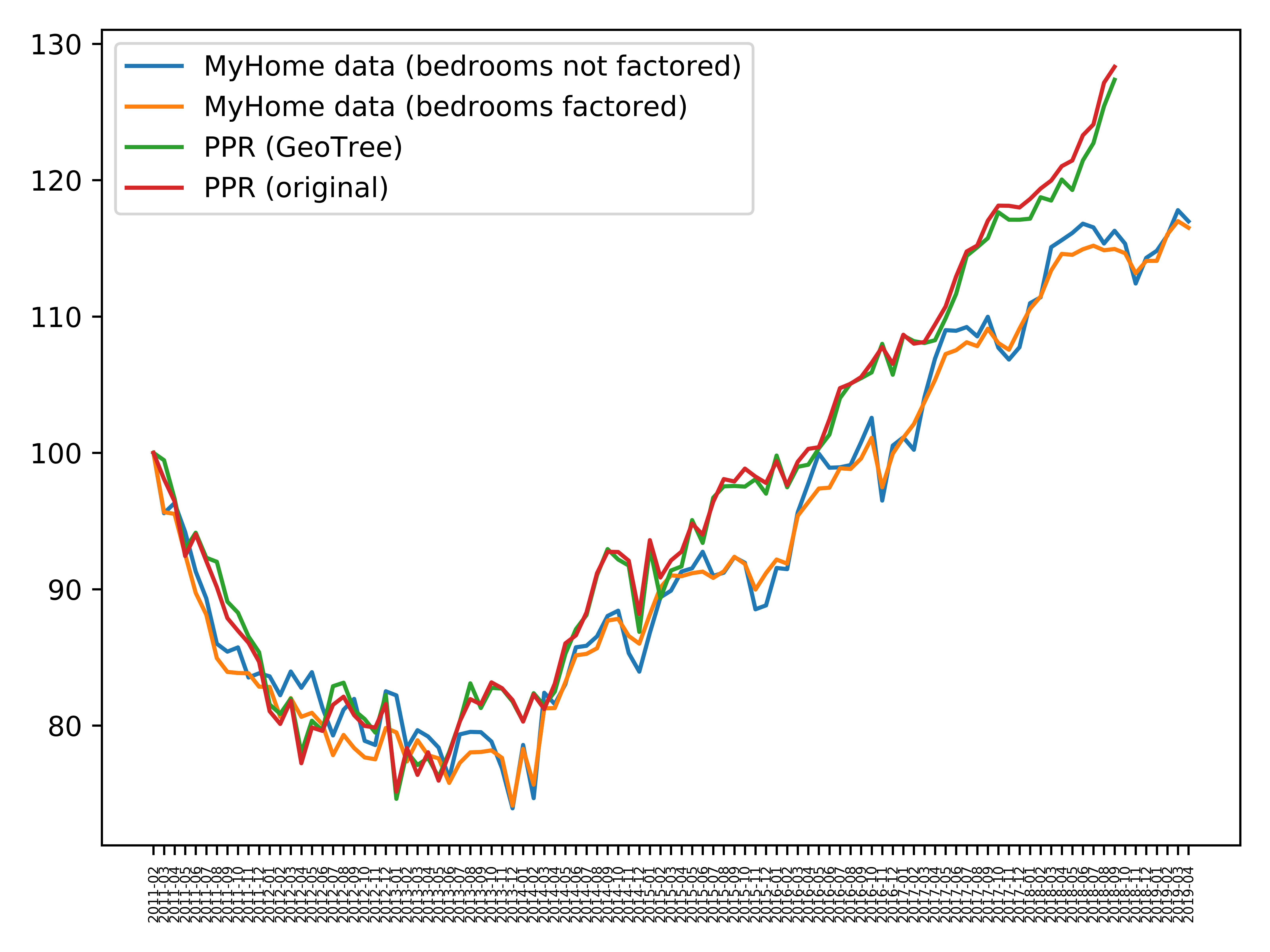}
			\caption{Comparison of index on PPR and MyHome data sets, from 02-2011 to 03-2019 [data limited to 09-2018 for PPR]}
			\label{fig: ppr_v_myhome_graph}
		\end{figure*}

		If we compare the MyHome results in isolation, we can clearly observe that the addition of bedroom matching makes a very significant impact on the index performance. While the trend of each graph is observably similar, \autoref{fig: ppr_v_myhome_graph} demonstrates that month to month changes are less erratic and appear less prone to large, spontaneous dips. Considering the smoothness metrics, the introduction of bedroom factoring generates a decrease of 26.8\% in the standard deviation of the differences and a decrease of approximately 48.4\% in the MSM. These results show a clear improvement by tightening the accuracy of property matching and are promising for the potential future inclusion of additional parameters such as bedroom matching should such data become available.
		
		\autoref{fig: ppr_v_myhome_graph} corresponds with the results of these metrics, with the \textit{MyHome data (bedrooms factored)} index appearing the smoothest time series of the four which are compared. It is important to note that the PPR data is based upon actual sale prices, while the MyHome data is based on listed asking prices of properties which are up for sale and as such, may produce somewhat different results.
		
		It is a well known fact that properties sell extremely well in spring and towards the end of the year, the former being the most popular period for property sales. Furthermore, the months towards late summer and shortly after tend to be the least busy periods in the year for selling property \cite{SpringANDeoyTOsell}. These phenomena can be observed in \autoref{fig: ppr_v_myhome_graph} where there is a dramatic increase in the listed asking prices of properties in the spring months and towards the end of each year, while the less popular months tend to experience a slump in price movement. As such, the two PPR graphs and the MyHome data (bedrooms not factored) graph are following more or less the same trend in price action and their graphs tend to meet often, however, the majority of the price action in the MyHome data graphs tends to wait for the popular selling months. The PPR graph does not experience these phenomena as selling property can be a long, protracted process and due to a myriad of factors such as price bidding, paperwork, legal hurdles, mortgage applications and delays in reporting, final sale notifications can happen outside of the time period in which the sale price is agreed between buyer and seller.
		
				\begin{table*}
			\caption{Index Comparison Statistics}
			\centering
			\newcolumntype{C}[1]{%
				>{\vbox to 4ex\bgroup\vfill\centering\arraybackslash}%
				p{#1}%
				<{\vskip-\baselineskip\vfill\egroup}}  
			\newcolumntype{?}[1]{!{\vrule width #1}}
			
			\resizebox{0.8\linewidth}{!}{
				\begin{tabular}{ | C{8em} ?{0.25em} C{5.5em} | C{5.5em} | C{5.5em} | }
					\hline Algorithm & St. Dev & St. Dev of Differences & MSM\\
					\hline \hline
					PPR (original) & 16.524 & 2.191 & 23.30 \\
					\hline
					PPR (GeoTree) & 16.378 & 2.518 & 29.78 \\
					\hline
					MyHome (without bedrooms) & 12.898 & 2.209 & 18.91 \\
					\hline
					MyHome (with bedrooms) &12.985 & 1.617 & 9.75 \\
					\hline
			\end{tabular}}
			\label{tab:index_stats}
		\end{table*}
		
		\section{Conclusion}
		
		The introduction of bedroom factoring as an additional parameter in the pairing of nearby properties has been shown to have a profound impact on the smoothness of the mix-adjusted median property price index, which was already shown to outperform a popularly used implementation of the hedonic regression model. This improvement was made possible due to the acquisition of a richer data set and the development of the GeoTree structure, which greatly increased the performance of the algorithm. There is future potential for the introduction of further property characteristics (such as the number of bedrooms, property type etc.) in the proximity matching part of the algorithm, should such data be acquired. 
		
		Furthermore, the design of the data structure used ensures that minimal computational complexity is added when considering the technical implementation of this algorithmic adjustment. As a result of this, the index can be computed quickly enough that it would be possible to have real-time updates (eg. up to every 5 minutes) to the price index, if a sufficiently rich stream of continuous data was available to the algorithm. Large property listing websites, such as Zillow, likely have enough \textit{live}, incoming data that such an index would be feasible to compute at this frequency, however, this volume of data is not publicly available for testing.

		\bibliographystyle{IEEEtran}
	\bibliography{IEEEabrv,references}
\end{document}